\documentclass{article}

\usepackage{microtype}
\usepackage{graphicx}
\usepackage{subfigure}
\usepackage{booktabs} % for professional tables

\usepackage{hyperref}

\usepackage[accepted]{icml2025}

\usepackage{amsmath}
\usepackage{amssymb}
\usepackage{mathtools}
\usepackage{amsthm}
\usepackage[capitalize,noabbrev]{cleveref}
\usepackage{lipsum}
\usepackage{stfloats} 
\theoremstyle{plain}

\theoremstyle{definition}

\theoremstyle{remark}

\usepackage{makecell}
\usepackage{mathrsfs}
\usepackage{algorithm}
\usepackage{algorithmic}
\usepackage{multirow}
\usepackage{graphicx}

\usepackage[textsize=tiny]{todonotes}

\icmltitlerunning{}

\begin{document}

\onecolumn
\icmltitle{A database of upper limb surface electromyogram signals from demographically diverse individuals}



\icmlsetsymbol{equal}{*}

\begin{icmlauthorlist}
\icmlauthor{Harshavardhana T. Gowda}{yyy}
\icmlauthor{Neha Kaul}{zzz}
\icmlauthor{Carlos Carrasco}{zzz}
\icmlauthor{Marcus Battraw}{www}
\icmlauthor{Safa Amer}{zzz}
\icmlauthor{Saniya Kotwal}{zzz}
\icmlauthor{Selena Lam}{zzz}
\icmlauthor{Zachary McNaughton}{zzz}
\icmlauthor{Ferdous Rahimi}{zzz}
\icmlauthor{Sana Shehabi}{zzz}
\icmlauthor{Jonathon Schofield}{ggg}
\icmlauthor{Lee M. Miller}{zzz,ccc,vvv,y}
%\icmlauthor{Firstname3 Lastname3}{comp}
%\icmlauthor{Firstname4 Lastname4}{sch}
%\icmlauthor{Firstname5 Lastname5}{yyy}
%\icmlauthor{Firstname6 Lastname6}{sch,yyy,comp}
%\icmlauthor{Firstname7 Lastname7}{comp}
%\icmlauthor{Firstname8 Lastname8}{sch}
%\icmlauthor{Firstname8 Lastname8}{yyy,comp}
\end{icmlauthorlist}

\icmlaffiliation{yyy}{Department of Electrical and Computer Engineering, University of California, Davis}
\icmlaffiliation{zzz}{Center for Brain and Mind, University of California, Davis}
\icmlaffiliation{ccc}{Department of Neurobiology, Physiology, and Behavior, University of California, Davis}
\icmlaffiliation{vvv}{Department of Otolaryngology/Head and Neck Surgery, University of California, Davis}
\icmlaffiliation{www}{Department of Mechanical and Mechatronic Engineering and Advanced Manufacturing, California State University, Chico}
\icmlaffiliation{ggg}{Department of Mechanical and Aerospace Engineering, University of California, Davis}
\icmlaffiliation{y}{\textcolor{blue}{A version of this article has been published in the {\em Scientific Data. 12.1 (2025): 517}}}
%\icmlaffiliation{comp}{Company Name, Location, Country}
%\icmlaffiliation{sch}{School of ZZZ, Institute of WWW, Location, Country}

\icmlcorrespondingauthor{Harshavardhana T. Gowda}{tgharshavardhana@gmail.com}
%\icmlcorrespondingauthor{Firstname2 Lastname2}{first2.last2@www.uk}

\icmlkeywords{Machine Learning, ICML}

\vskip 0.3in



\printAffiliationsAndNotice{}  % leave blank if no need to mention equal contribution
\begin{abstract}
Upper limb based neuromuscular interfaces aim to provide a seamless way for humans to interact with technology. Among noninvasive interfaces, surface electromyogram (EMG) signals hold significant promise. However, their sensitivity to physiological and anatomical factors remains poorly understood, raising questions about how these factors influence gesture decoding across individuals or groups. To facilitate the study of signal distribution shifts across individuals or groups of individuals, we present a dataset of upper limb EMG signals and physiological measures from 91 demographically diverse adults. Participants were selected to represent a range of ages (18 to 92 years) and body mass indices (healthy, overweight, and obese). The dataset also includes measures such as skin hydration and elasticity, which may affect EMG signals. This dataset provides a basis to study demographic confounds in EMG signals and serves as a benchmark to test the development of fair and unbiased algorithms that enable accurate hand gesture decoding across demographically diverse subjects. Additionally, we validate the quality of the collected data using state-of-the-art gesture decoding techniques.
\end{abstract}

%%%%%%%%%%%%%%%%%%%%%%%%%%%%%%%%%%%%%%%%%%%%%%%%%%%%%%%%%%%%%%%%%%%%%%%%%%%%%%%%%%%%%%%%%%%%%%%%%%%%%%%%%%%%%%%%%%%%%%%%%%%%%%%%%%%%%%%
\section{Background \& Summary}\label{sec1}
Surface electromyogram (EMG) signals, measured non-invasively at the skin surface, capture muscle activation patterns associated with corresponding physical movements. Hand gestures serve as a prominent modality of human communication and interaction, making upper-limb EMG-based neural interfaces a promising tool for various applications, including hand-gesture-based computer interfaces \cite{salteremg2pose}, handwriting decoding \cite{ctrl2024generic}, keyboard typing \cite{sivakumaremg2qwerty}, and augmenting supernumerary limbs or fingers. However, a significant challenge in deploying such neural interfaces at scale lies in the variability of EMG signals across individuals due to anatomical and physiological differences. At present, our understanding of how demographic factors influence EMG signals - and, consequently, decoding performance - remains limited. To address these critical questions, we introduce a dataset capturing EMG data across diverse demographic groups, enabling systematic investigation into these issues.

Recent advancements have led to the development of easily wearable EMG interfaces, as described by \citeauthor{ctrl2024generic} These interfaces are designed to facilitate widespread use in human-computer and human-robot interaction, with the goal of democratizing access to EMG technology and transforming paradigms of human-computer interaction. However, EMG signals are significantly affected by distributional shifts across individuals due to inter-subject differences in neural drive and muscle properties \cite{farina2014extraction, farina2004extraction}. Factors such as subcutaneous fat thickness, the spatial distribution of muscle fibers, variations in muscle fiber conduction velocity \cite{farina2014extraction}, and contextual elements like electrode placement contribute to this variability. Additionally, neural properties, such as the discharge characteristics of the neural drive, further exacerbate the inter-individual differences in EMG signals \cite{farina2014extraction}.
These differences underscore the importance of collecting data from demographically diverse populations to develop fair and inclusive machine learning models for muscle activity detection so that such models can account for inter-individual variability to ensure equitable performance and robust applicability across diverse user groups.

Machine learning models trained on imbalanced datasets that exclude certain demographic groups often produce erroneous classifications on the underrepresented groups. For example, early facial analysis benchmarks, such as IJB-A \cite{klare2015pushing}, were predominantly composed of lighter-skinned subjects, accounting for nearly 80\% of the dataset, as highlighted by \citeauthor{buolamwini2018gender}. Similarly, many private datasets used for training machine learning models had skewed demographic representation, and consequently some facial detection software failed to recognize dark-skinned faces \cite{buolamwini2018robot}.  Moreover, commercially available gender classification systems showed the highest misclassification rates for the underrepresented darker-skinned females, with error rates approaching 35\%, compared to just 0.8\% for lighter-skinned males \cite{buolamwini2018gender}.
Furthermore, biases present in datasets used to train language models, such as word2vec \cite{mikolov2013efficient}, are reflected in the word embedding spaces. For instance, \citeauthor{bolukbasi2016man} demonstrated this issue using an analogy generator trained on word embeddings. The analogy \textsc{man:computer programmer::woman:X} was completed with \textsc{X = homemaker}, reinforcing harmful societal stereotypes. Building on these observations, and to ensure that electromyography-based interfaces function equitably across diverse demographic groups, we have developed a demographically inclusive dataset.

Age-related decline in muscle strength has been well-documented \cite{Goodpaster2006}.
Additionally, obesity-induced attenuation of calcium signaling, mediated through its effects on calcineurin, adiponectin, and actinin, disrupts excitation-contraction and excitation-transcription coupling in myocytes. These molecular alterations adversely affect muscle contractile function, leading to reductions in both the force generated per unit of muscle cross-sectional area and the power produced per unit of muscle mass \cite{tallis2018effects}. Consequently, there is a critical need to investigate the signal distribution shifts in surface electromyography (EMG) data in older individuals and those with high body mass indices (BMI). To address this, our dataset includes participants spanning a range of age groups and BMI categories. In addition, we measure skin elasticity and hydration levels, as these factors may influence EMG recordings by altering skin conductivity. While there are several existing open-access EMG datasets \cite{amma2015advancing, atzori2014electromyography, du2017surface, palermo2017repeatability, jiang2021open, sivakumaremg2qwerty}, their primary focus has been on the clinical feasibility of EMG-based neuroprostheses and applications in human-computer interaction (HCI) and they do not aim to create demographically inclusive corpora. \citeauthor{salteremg2pose} provide EMG data with limited demographic diversity with all subjects under the age of 60 years; whereas, our dataset includes 52 individuals above the age of 60, with 23 of them having high BMI - providing a new opportunity to explore if older adults and older adults with high BMI can effectively use EMG devices described by \citeauthor{ctrl2024generic} A part of the data presented here has been used to demonstrate the geometrical structure of the EMG signals in \citeauthor{gowda2024topology}. We provide a comparative analysis of our dataset with previous datasets in table \ref{tab:1}.

\begin{table}[h!]
\centering
\begin{tabular}{l l l l}
\hline
\textbf{Dataset} & \textbf{\# subjects} & \textbf{Application} & \textbf{Demographically diverse?} \\ \hline
\citeauthor{amma2015advancing} & 5 & HCI & No \\ \hline
\citeauthor{atzori2014electromyography} & 78 & HCI, Neuroprostheses & No \\ \hline
\citeauthor{du2017surface} & 23 & HCI & No \\ \hline
\citeauthor{palermo2017repeatability} & 10 & Neuroprostheses & No \\ \hline
\citeauthor{jiang2021open} & 20 & HCI, Neuroprostheses & No\\ \hline
\citeauthor{malevsevic2021database} & 20 & HCI & No \\\hline
\citeauthor{sivakumaremg2qwerty} & 108 & HCI & Unknown \\ \hline
\citeauthor{salteremg2pose} & 193 & HCI & Limited (all individuals under 60 years) \\ \hline
\textbf{Ours} & \textbf{91} & \textbf{HCI} & \textbf{Yes} \\ \hline
\end{tabular}
\caption{Comparison with prior datasets.}
\label{tab:1}
\end{table}

\begin{figure}[h!]
    \centering
    \includegraphics[width=0.7\textwidth]{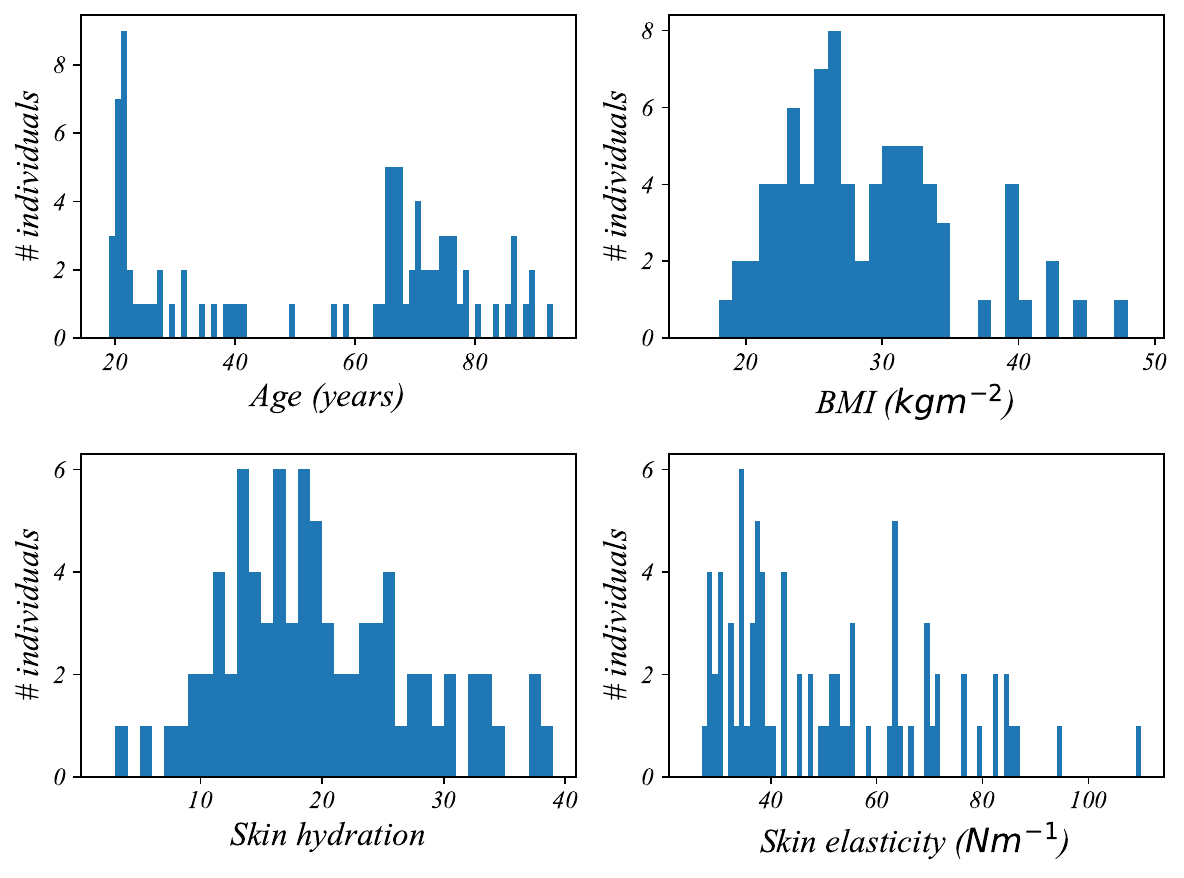}
    \caption{Histograms of age, BMI, skin hydration, and skin elasticity.}
    \label{fig:distributionGroups}
\end{figure}

To the best of our knowledge, this is the only dataset that provides EMG data specifically from older adults and individuals with high BMI. The histogram distribution of demographic groups is illustrated in figure \ref{fig:distributionGroups}. Key features of the dataset include:

\begin{list}{}{\setlength{\leftmargin}{3.5em} \setlength{\rightmargin}{0pt} \setlength{\itemindent}{8pt} \setlength{\labelwidth}{2.5em} \setlength{\labelsep}{0.5em}}
    \item[\textcircled{\footnotesize 1} \textbf{Intersectional demographic analysis:}] The dataset supports an intersectional evaluation of EMG signals, enabling studies on combinations such as `older + high BMI', `younger + high BMI', `older + high skin hydration', and more.

    \item[\textcircled{\footnotesize 2} \textbf{Comprehensive trial collection:}] The dataset comprises 32,760 hand gesture trials from 91 participants, with each subject contributing 360 trials (10 distinct hand gestures, each repeated 36 times).

    \item[\textcircled{\footnotesize 3} \textbf{Natural gesture performance:}] Participants performed gestures in a natural, unconstrained manner, reflecting typical hand movements. This approach contrasts with prior works, such as \citeauthor{malevsevic2021database} and \citeauthor{atzori2014electromyography}, where participants' hands were restricted using force measurement devices. As a result, our dataset offers a more realistic representation of natural hand gestures.

    \item[\textcircled{\footnotesize 4} \textbf{Comprehensive signal coverage:}] The dataset includes EMG signals collected from both the wrist and forearm regions.
\end{list}

Additionally, in the technical validation section, we show that recognition of different hand gestures is possible across individuals from all demographics using state-of-the-art methods.

%%%%%%%%%%%%%%%%%%%%%%%%%%%%%%%%%%%%%%%%%%%%%%%%%%%%%%%%%%%%%%%%%%%%%%%%%%%%%%%%%%%%%%%%%%%%%%%%%%%%%%%%%%%%%%%%%%%%%%%%%%%

\section{Methods}

\subsection{Subjects and ethical requirements}
A total of 91 subjects (age range: 18-92 years; mean age: 53.53 years; age standard deviation: 24.37 years; 37 males and 54 females) participated in our study. Research was conducted in accordance with the principles embodied in the Declaration of Helsinki and in accordance with the University of California Davis Institutional Review Board Administration protocol {2078695-1}. All participants provided written informed consent. Consent was also given for publication of the deidentified data by all participants. Participants were healthy volunteers and were selected from any gender and all ethnic and racial groups. Subjects were aged 18 or above, were able to fully understand spoken and written English, and were capable of following task instructions. Subjects had no skin conditions or wounds where electrodes were placed. Subjects were excluded if they had uncorrected vision problems or neuromotor disorders that prevented them from making hand gestures. Children, adults who were unable to consent, and prisoners were not included in the experiments. 

\subsection{Acquisition setup}
EMG data were collected using Delsys Trigno double-differential electrodes (\href{www.delsys.com}{Delsys, Inc}) and an NI USB-6210 multifunction I/O device (\href{www.ni.com}{National Instruments Corporation} - 16 inputs, 16-bit, 250 kS/s) at sampling rates of either 2000 Hz or 2148 Hz. The Delsys electrodes wirelessly transmitted data to a base station, which subsequently relayed the information to a computer via the NI USB-6210 system using a USB connection.  

A custom graphical user interface (GUI) was designed to display hand gestures on a screen, facilitating participant interaction. Subjects, seated comfortably with their dominant forearm resting on an elevated platform, performed gestures displayed on the screen. Participants were allowed to choose and adjust their resting position throughout the experiment. Each gesture was displayed for 2 seconds, during which participants performed the gesture, followed by a 2-second resting period indicated by a blank screen.  

The experiment consisted of six sessions, each comprising 60 trials, with six repetitions of 10 distinct gestures per session. Gesture order within each session was pseudorandomly generated to introduce variability in gesture performance and avoid repetitive or overly consistent movements. This approach ensured a more natural representation of gestures. In total, each participant completed 360 trials. Synchronization between the GUI and data acquisition was achieved using ZeroMQ sockets (\href{https://zeromq.org/socket-api/}{ZeroMQ}) and Lab Streaming Layer (\href{https://github.com/sccn/labstreaminglayer}{Lab Streaming Layer}) in Python. Both EMG data and event markers were synced to the computer's master clock.  

\textbf{Demographic and physiological data:}  
we collected self-reported demographic information, including age, height, and weight, along with physiological measures such as subcutaneous fat, hair density, skin elasticity, and skin moisture on the forearm.  

\begin{list}{}{\setlength{\leftmargin}{3.5em} \setlength{\rightmargin}{0pt} \setlength{\itemindent}{8pt} \setlength{\labelwidth}{2.5em} \setlength{\labelsep}{0.5em}}
    \item[\textcircled{\footnotesize 1} \textbf{Skin elasticity:}] measured using the Delfin Elastimeter (\href{https://delfintech.com/products/elastimeter/}{Delfin Elastimeter}), which utilizes an indenter that is briefly pressed onto the skin. When an external load is applied, the skin resists deformation, and its response under a short-term load reflects its instantaneous elastic properties. The presented values are averaged over five successive measurements taken at the same skin site.
    
    \item[\textcircled{\footnotesize 2} \textbf{Skin moisture:}] assessed using the Delfin MoistureMeterSC (\href{https://delfintech.com/products/moisturemetersc/}{Delfin MoistureMeterSC}). The probe head, the skin surface, and the deeper skin layers together form a structure similar to an electrical capacitor. The measured capacitance is proportional to the water content in the surface layer of the skin. A higher measured value indicates a higher moisture content in the stratum corneum.
    
    \item[\textcircled{\footnotesize 3} \textbf{Hair density:}] evaluated at four specific locations on the arm - anterior wrist, posterior wrist, anterior upper arm, and posterior upper arm - using the Aram Huvis API 202 device (\href{https://www.aramhuvis.com/en/product/api-202}{Aram Huvis API 202}). This device utilizes high-resolution imaging and specialized software to analyze hair density by capturing magnified images of the skin surface and automatically detecting individual hair strands. The measurement process involves placing the device on each designated skin site, ensuring consistent alignment to obtain accurate readings.

    \item[\textcircled{\footnotesize 4} \textbf{Subcutaneous fat:}] measured at two locations - the posterior wrist and the posterior forearm - using a MEDCA body fat caliper. This device estimates subcutaneous fat thickness by gently pinching the skin and underlying fat layer at each site with a calibrated caliper. The measurement is taken by applying consistent pressure to ensure accurate and reproducible readings. These values provide an indirect assessment of overall body fat distribution.
\end{list}

The distribution of age, BMI, skin elasticity, and hydration is shown in figure \ref{fig:distributionGroups}. However, subcutaneous fat and hair density measurements did not exhibit significant variability across the population, likely due to limitations in device precision.

\subsection{Acquisition protocol}
\begin{figure}[h!]
	\centering
	\includegraphics[width=6cm]{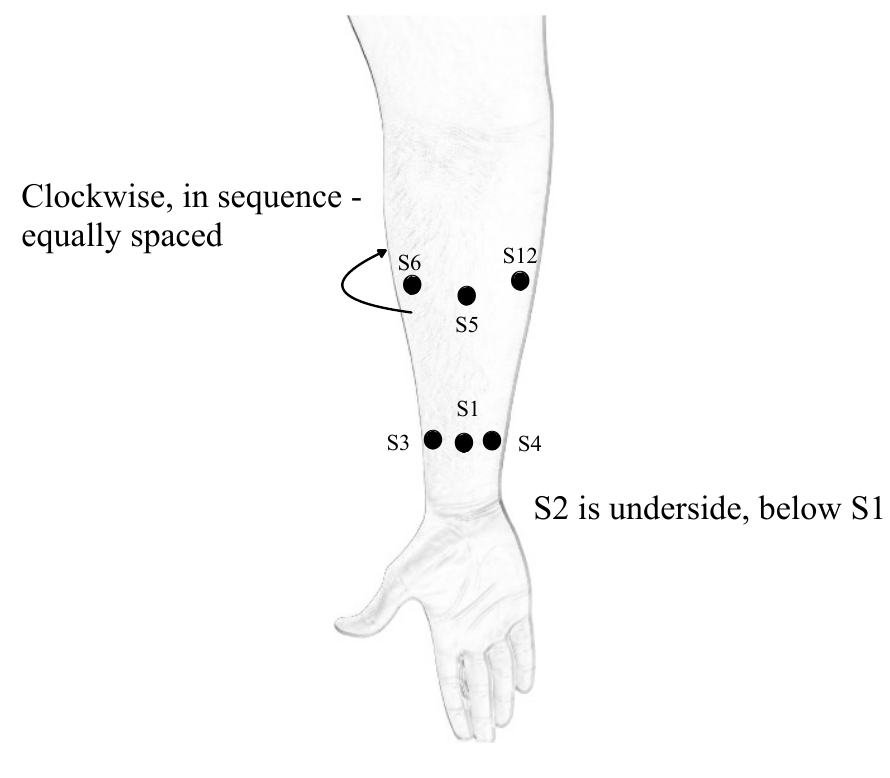}
	\caption{\footnotesize Sensors are named with a prefix {\em S} followed by the sensor number. 4 sensors are placed equally spaced around the wrist joint ({\em S}1, {\em S}3, {\em S}4 are shown. {\em S}2 is on the underside, below {\em S}1). Sensors {\em S}5 to {\em S}12 are placed equally spaced around the belly of the forearm. Sensors {\em S}5 and {\em S}6 are shown. The rest of the sensors are equally spaced around the forearm in the clockwise direction so that {\em S}12 is adjacent to {\em S}5 as shown.}
	\label{fig:Arm}
\end{figure}

\begin{figure}[h!]
	\centering
	\includegraphics[width=8cm, angle = -90]{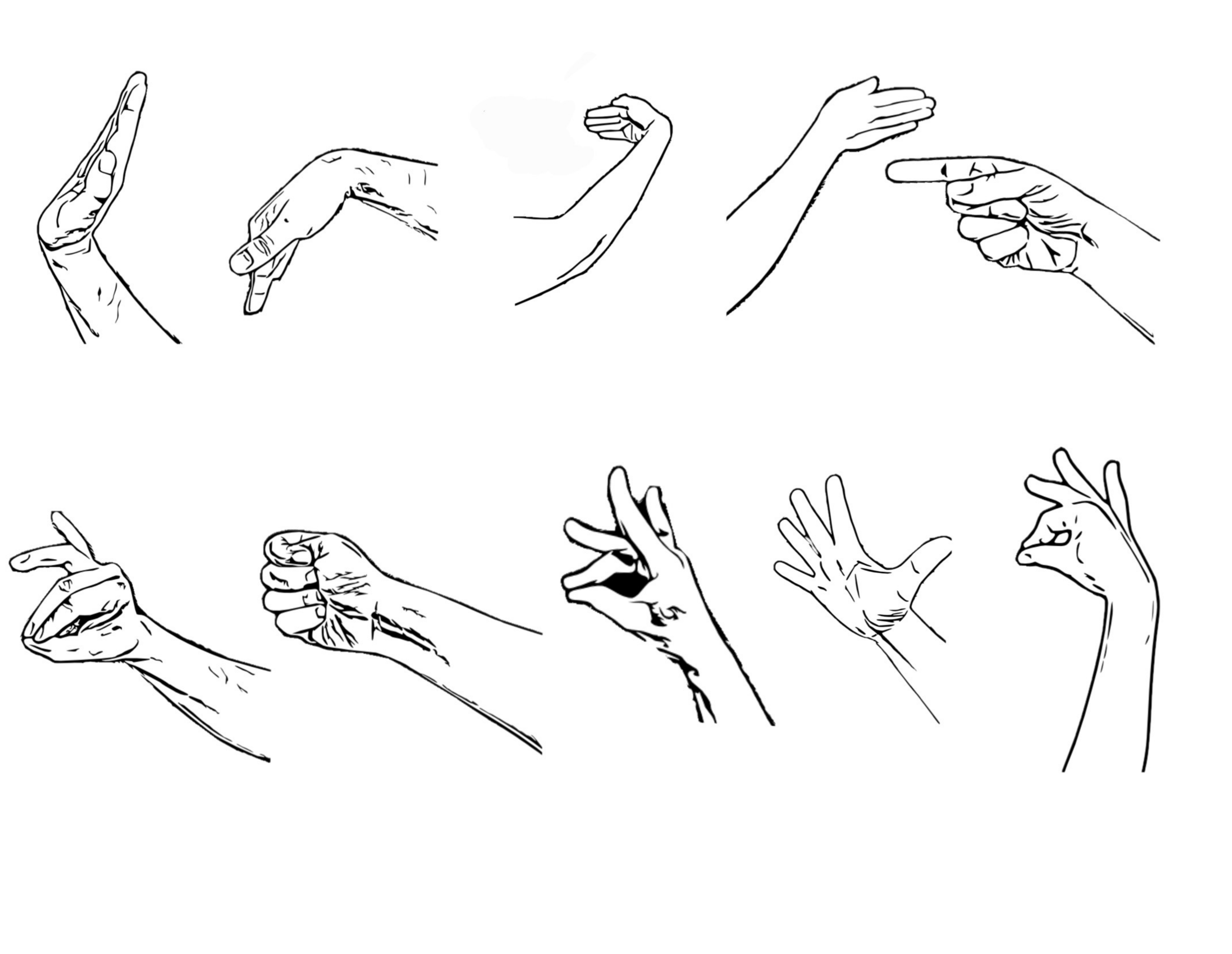}
	\caption{\footnotesize Ten gestures included in the experiment. From top-left: up, down, left, right, index point, two finger pinch, power grasp, middle finger pinch, splay, index finger pinch. Image is reproduced from \citeauthor{gowda2024topology}.}
	\label{fig:Gestures}
\end{figure}

 Forearm EMG was collected from the upper limb using 12 electrodes. Eight electrodes were placed equally spaced around the main belly of the forearm muscles below the elbow at
approximately 1/3 the distance from elbow to wrist. Four electrodes were placed equally spaced around the wrist joint (figure \ref{fig:Arm}). Each subject performed ten different hand gestures (figure \ref{fig:Gestures}), with each gesture performed thirty-six times. Ten gestures are labeled as follows: 1 - Down, 2 - Index finger pinch, 3 - Left, 4 - Middle finger pinch, 5 - Index point, 6 - Power grasp, 7 - Right, 8 - Two finger pinch, 9 - Up, 10 - Splay.

Before signal acquisition, participants were briefed on the experimental protocol. They were seated comfortably on a chair with their dominant arm resting on an elevated platform. Participants were instructed to perform the gestures naturally, mimicking how they would execute these movements in everyday situations.

Each gesture was performed during a 2-second window, in course of which an image of the gesture was displayed on a graphical user interface (GUI). The 10 selected gestures were designed to represent common actions used for interacting with computers. Gestures in the cardinal directions (up, down, left, and right) were included to facilitate screen navigation, while pinch gestures were intended for actions like zooming in, zooming out, and making selections.

\subsection{Data preprocessing}
Minimal signal processing was applied to the data prior to its release on the public repository \cite{miller_gowda_2025}. These preprocessing steps included time synchronization and signal segmentation. The continuous EMG signal stream, acquired from the Delsys system through a Python input socket, was time-synchronized to the master clock using Lab Streaming Layer. Signals were segmented into 2-second trials corresponding to the duration of each gesture display on the GUI.

The data collection environment was carefully controlled to eliminate AC electrical interference. Additionally, the Delsys system applied built-in hardware filtering to the EMG signals, restricting the frequency range to 20-450 Hz.
%%%%%%%%%%%%%%%%%%%%%%%%%%%%%%%%%%%%%%%%%%%%%%%%%%%%%%%%%%%%%%%%%%%%%%%%%%%%%%%%%%%%%%%%%%%%%%%%%%%%%%%%%%%%%%%%%%%%%%%%
\section{Data Records}
The data presented here can be downloaded from \href{https://osf.io/3kzcb/}{OSF} \cite{miller_gowda_2025}. Below, we describe the contents of the provided files.

The dataset is publicly available in \textsc{pickle} (\textsc{.pkl}) format, structured as a Python \textsc{dictionary} with four primary keys:

   \textcircled{\footnotesize 1} \textbf{\textsc{EMG}}: Contains the EMG data as a \textsc{numpy} array of shape \textsc{(360, 12, t)}, where \textsc{t} represents the number of time points corresponding to the 2-second gesture duration (either 4000 or 4296 samples, depending on whether the sampling frequency was 2000 Hz or 2148 Hz). The dataset includes 360 trials from each subject, distributed across six sessions with 60 trials per session. EMG was recorded from 12 electrodes (refer to figure \ref{fig:Arm} for electrode placement).

    \textcircled{\footnotesize 2} \textbf{\textsc{Labels}}: A \textsc{numpy} array of shape \textsc{(360,)} containing gesture labels corresponding to each trial in the \textsc{EMG} data.

    \textcircled{\footnotesize 3} \textbf{\textsc{Frequency}}: An \textsc{integer} specifying the sampling frequency of the EMG data (2000 Hz or 2148 Hz).

    \textcircled{\footnotesize 4} \textbf{\textsc{Physiology}}: A nested \textsc{dictionary} that includes demographic and physiological data for each subject:
    \begin{list}{}{\setlength{\leftmargin}{3.5em} \setlength{\rightmargin}{0pt} \setlength{\itemindent}{8pt} \setlength{\labelwidth}{2.5em} \setlength{\labelsep}{0.5em}}
        \item[\textcircled{\tiny 4a} \textbf{\textsc{Age}}]: An \textsc{integer} representing the self-reported age in years.
        
      \item[\textcircled{\tiny 4b} \textbf{\textsc{Height}}]: A \textsc{float} representing the self-reported height in centimeters.
       
       \item[\textcircled{\tiny 4c} \textbf{\textsc{Weight}}]: A \textsc{float} representing the self-reported weight in kilograms.
       
       \item[\textcircled{\tiny 4d} \textbf{\textsc{Skin hydration}}]: A \textsc{float} indicating the moisture level of the forearm skin. For details on measurement units, range, and calibration, refer to \href{https://delfintech.com/products/moisturemetersc/}{Delfin MoisturemeterSC}.
       
     \item[\textcircled{\tiny 4e} \textbf{\textsc{Skin elasticity}}]: A \textsc{float} representing the elasticity of the forearm skin in Nm\textsuperscript{-1}.
       
      \item[\textcircled{\tiny 4f} \textbf{\textsc{Subcutaneous fat}}]: A \textsc{list} with subcutaneous fat measurements (in millimeters) taken from four locations on the arm: forearm anterior, forearm posterior, wrist anterior, and wrist posterior (presented in this order).
       
     \item[\textcircled{\tiny 4g} \textbf{\textsc{Hair density}}]: A \textsc{list} with hair density measurements (in hairs/cm\textsuperscript{2}) taken from two locations: forearm anterior and wrist anterior (presented in this order).
      
     \item[\textcircled{\tiny 4h} \textbf{\textsc{Sex}}]: Self-reported biological sex: \textsc{F} (Female), \textsc{M} (Male), or \textsc{N} (Non-binary/Other).
      \end{list}
If a physiological measurement is unavailable for a subject, the corresponding field is assigned the placeholder value \textsc{None}.
%%%%%%%%%%%%%%%%%%%%%%%%%%%%%%%%%%%%%%%%%%%%%%%%%%%%%%%%%%%%%%%%%%%%%%%%%%%%%%%%%%%%%%%%%%%%%%%%%%%%%%%%%%%%%%%%%%%%%%%%%%%%%%%%%

\section{Technical Validation}
To evaluate the quality of the collected data, we perform hand gesture classification following the methods outlined by \citeauthor{gowda2024topology}. Their approach involves constructing covariance matrices from EMG signals and analyzing them on the manifold of symmetric positive definite (SPD) matrices. The non-Euclidean matrix embeddings corresponding to different hand gestures are naturally separable on the manifold of SPD matrices. Using these representations, they demonstrated high decoding accuracy with methods such as minimum distance to the mean (MDM), support vector machines (SVM), and unsupervised {\em k} - medoids clustering using Riemannian geodesic distance. Adopting their framework, we train these models for each individual separately and report the average decoding accuracy across all participants in table \ref{tab:2}.

\begin{table}[h!]
\centering
\begin{tabular}{l l l}
\hline
\textbf{MDM} & \textbf{SVM ($\gamma = 1$)} & \textbf{{\em k}-medoids}   \\ \hline
 0.769 $\pm$ 0.171 & 0.800 $\pm$ 0.170  & 0.639 $\pm$ 0.211 \\ \hline

\end{tabular}
\caption{Mean decoding accuracy across all 91 subjects using various methods given by \citeauthor{gowda2024topology}. Chance decoding accuracy is 0.1. We see that the obtained decoding accuracy are much higher than the chance levels, confirming the quality of the obtained signals. For supervised MDM and SVM methods, data from first 3 sessions are used for training and data from last 3 sessions are used for testing.}
\label{tab:2}
\end{table}

%%%%%%%%%%%%%%%%%%%%%%%%%%%%%%%%%%%%%%%%%%%%%%%%%%%%%%%%%%%%%%%%%%%%%%%%%%%%%%%%%%%%%%%%%%%%%%%%%%%%%%%%%%%%%%%%%%%%%%%%%%%%%%%%%%%%%

\subsection{Geometric structure of the EMG signals is insensitive to demographics}

We explore if the inherent geometric structure of the EMG signals as described by \citeauthor{gowda2024topology} is influenced by factors such as age, skin hydration, skin elasticity, and BMI. That is, we analyze if the geometric structure is pronounced (or unremarkable) in certain population groups. To access this, we make use of classification accuracy obtained using unsupervised {\em k} - medoids algorithm on the manifold as this algorithm encapsulates all the geometric structure of the data including how densely (or sparsely) the SPD matrices (belonging to a particular gesture) cluster on the manifold and how distinguishable (far apart) the clusters belonging to different gestures are and analyze the trends with respect to age, skin hydration, skin elasticity, and BMI. 

We calculate variance inflation factor (VIF) for age, skin hydration, skin elasticity, and BMI to verify that no significant multicollinearity exists between these factors. The values are summarized in table \ref{tab:VIF}. The VIF values are well below 5 and do not show any problematic multicollinearity.  
\begin{table}[h!]
	\centering
	\begin{tabular}{l l}
		\hline
		\textbf{Feature} & \textbf{VIF} \\ \hline
		Age        & 2.20 \\ \hline
		Skin hydration  & 1.21 \\ \hline
		Skin elasticity & 1.86 \\ \hline
		BMI        & 1.141 \\ \hline
	\end{tabular}
	\caption{Variance inflation factor (VIF) for features age, skin hydration, skin elasticity, and BMI.}
	\label{tab:VIF}
\end{table}

\begin{figure}[h!]
	\centering
	\includegraphics[width=10cm]{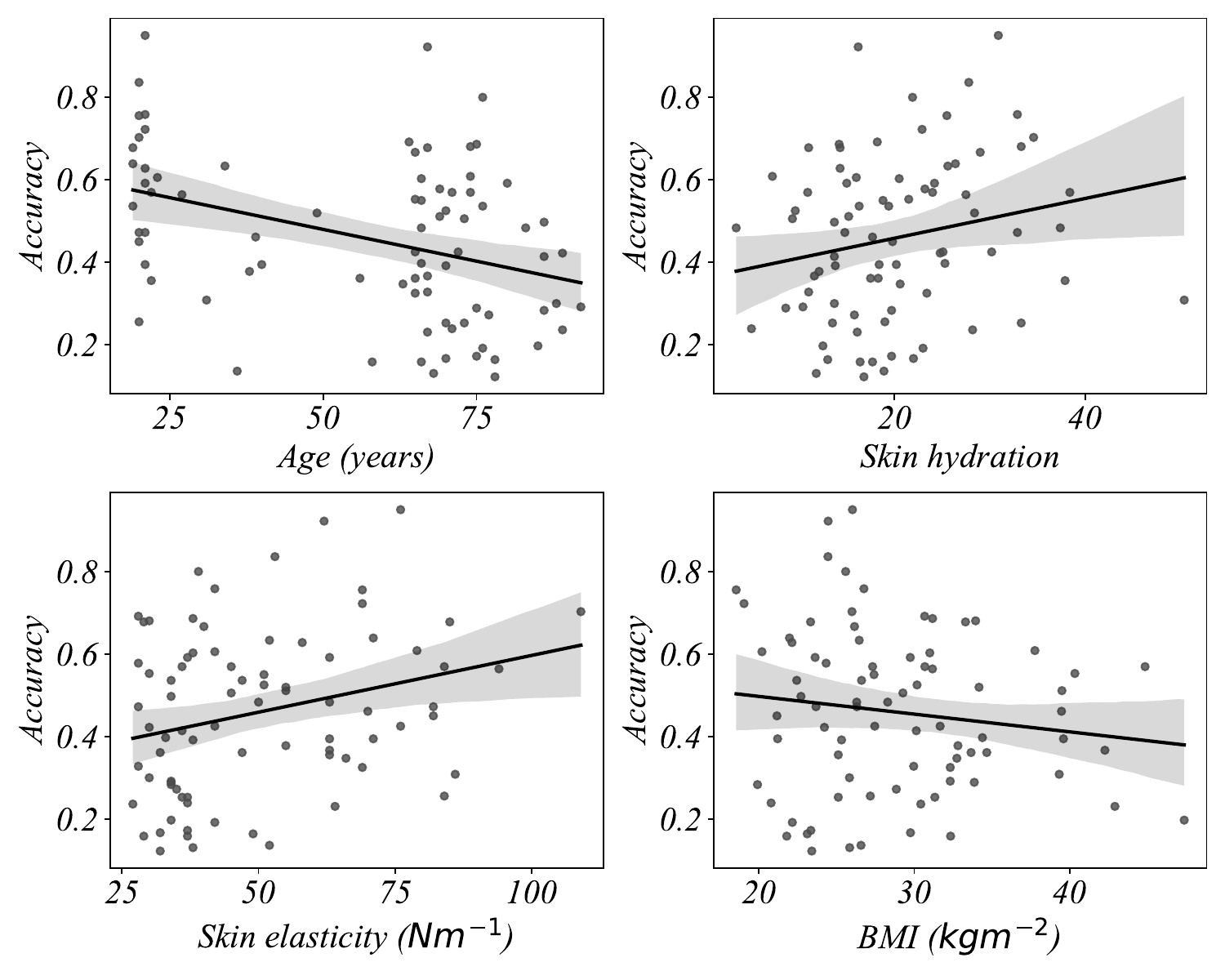}
	\caption{Regression plots of {\em k} - medoids classification accuracy with age, skin hydration, skin elasticity, and BMI for the frequency range 20 to 50 Hertz.}
	\label{fig:freq0Regression}
\end{figure}

\begin{figure}[h!]
	\centering
	\includegraphics[width=10cm]{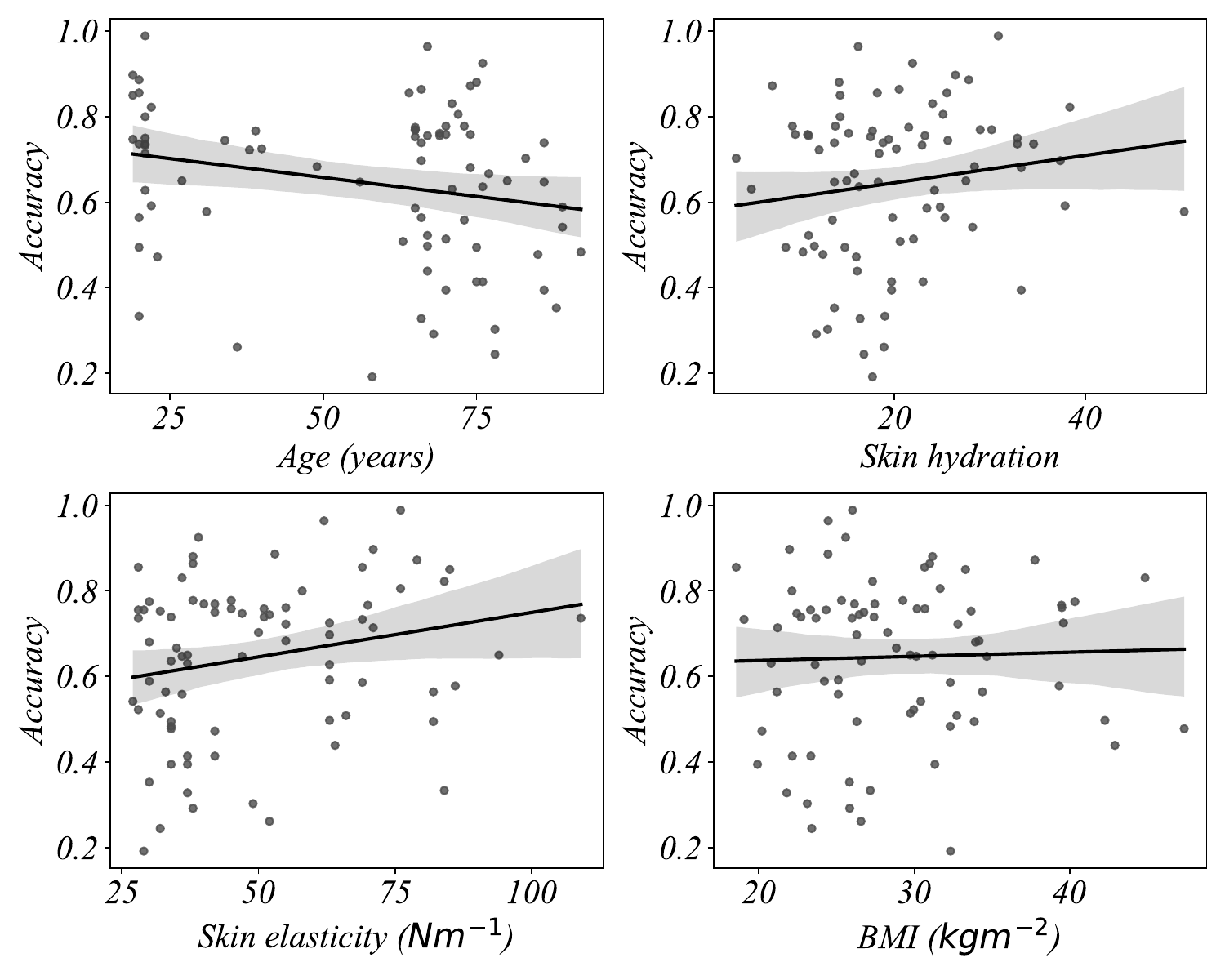}
	\caption{Regression plots of {\em k} - medoids classification accuracy with age, skin hydration, skin elasticity, and BMI for the frequency range 50 to 110 Hertz.}
	\label{fig:freq1Regression}
\end{figure}

\begin{figure}[h!]
	\centering
	\includegraphics[width=10cm]{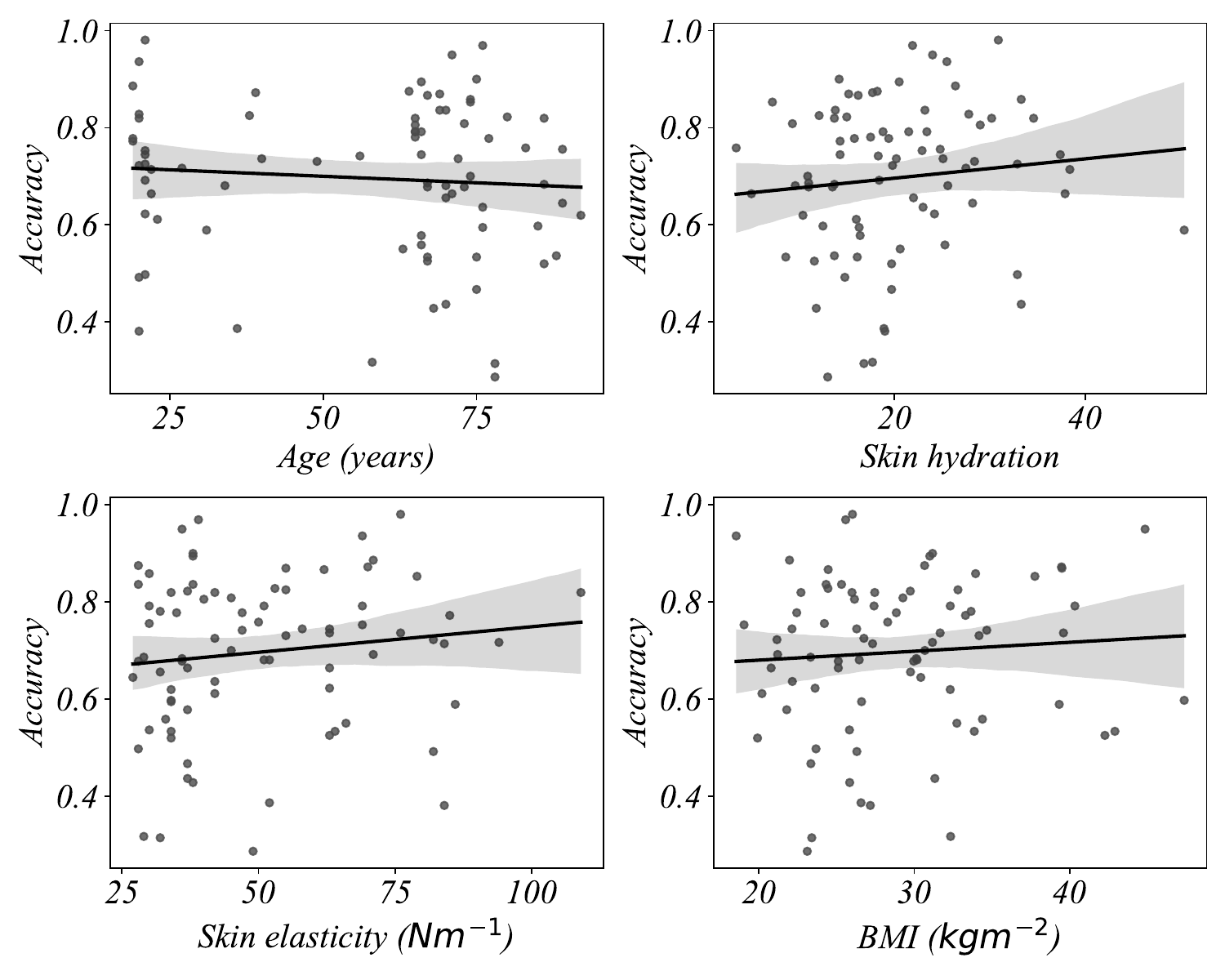}
	\caption{Regression plots of {\em k} - medoids classification accuracy with age, skin hydration, skin elasticity, and BMI for the frequency range 110 to 230 Hertz.}
	\label{fig:freq2Regression}
\end{figure}

\begin{figure}[h!]
	\centering
	\includegraphics[width=10cm]{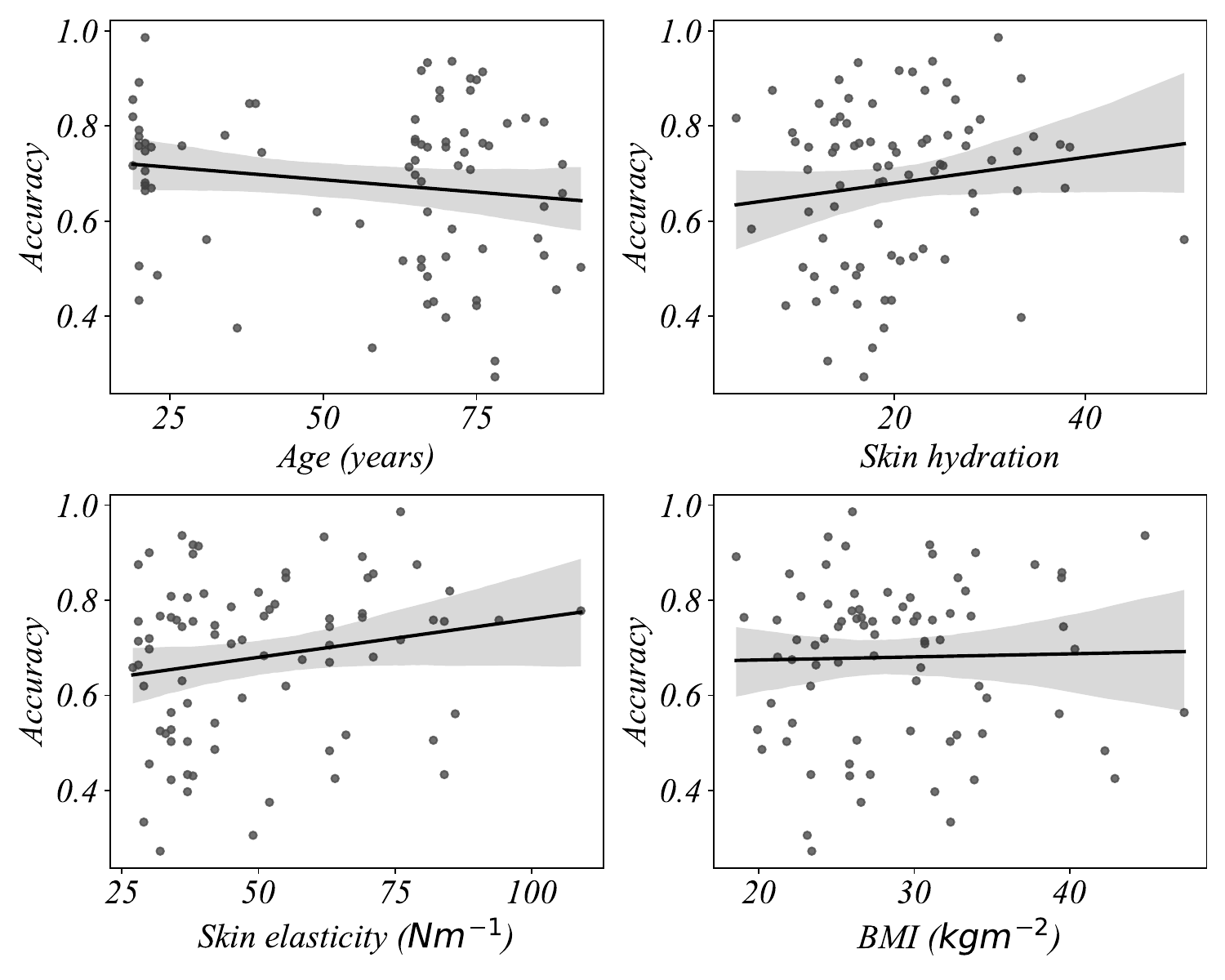}
	\caption{Regression plots of {\em k} - medoids classification accuracy with age, skin hydration, skin elasticity, and BMI for the frequency range 230 to 450 Hertz.}
	\label{fig:freq3Regression}
\end{figure}

We then perform multiple regression and find the linear relationship between the {\em k-}medoids decoding accuracy across various frequency ranges (dependent variable)\footnote{Raw EMG signals are bandpass filtered using third order Butterworth filter between the required frequencies.} and age, skin hydration, skin elasticity, and BMI (independent variables) using ordinary least square regression. 
The proportion of variance the models explain ({\em R}$^2$) are summarized in table \ref{tab:R2}. We observe that the $R^2$ values are relatively low indicating that the variables age, skin hydration, skin elasticity, and BMI do not substantially explain the variance in the decoding accuracy. That is, the geometry of the EMG signals is only slightly affected by these variables and higher frequencies are less confounded compared to lower frequencies. Also, as indicated by the $p$-values, only the model for frequency range 20 to 50 Hz is statistically significant. That is, at least one independent variable meaningfully explains the variability in the dependent variable in that frequency range. In all other frequency ranges, the models are not statistically significant (no independent variable meaningfully explains the variability in the dependent variable). Detailed regression model parameters are given in figures \ref{fig:freq0Regression}, \ref{fig:freq1Regression}, \ref{fig:freq2Regression}, and \ref{fig:freq3Regression} and tables \ref{tab:RT1}, \ref{tab:RT2}, \ref{tab:RT3}, and \ref{tab:RT4}.

A plausible explanation for the greater confounding of lower frequency bands by demographics is the potential interference from signal artifacts, which could arise from factors like motion and relative movement between the skin surface and electrodes. These movements can lead to changes in skin impedance, which in turn may distort the signal, especially at lower frequencies \cite{clancy2002sampling}. While this remains a hypothesis, a more detailed investigation into the sources of these artifacts and their impact on signal quality is needed to better understand the underlying mechanisms and confirm the role they play in these observations.
\begin{table}[h!]
	\centering
	\begin{tabular}{l l l}
		\hline
		Frequency & {\em R}$^2$ & {\em p}-value \\ \hline
		20 {\em to} 50 Hz        & 0.148 & 0.0164 \\ \hline
		50 {\em to} 110 Hz  & 0.069 & 0.245 \\ \hline
		110 {\em to} 230 Hz & 0.027 & 0.721 \\ \hline
		230 {\em to} 450 Hz        & 0.043 & 0.501 \\ \hline
	\end{tabular}
	\caption{Proportion of variance in decoding accuracy explained by age, skin hydration, skin elasticity, and BMI, with corresponding {\em p}-values. Signals are filtered using third order Butterworth bandpass filter.}
	\label{tab:R2}
\end{table}

\newpage

\begin{table}[h!]
	\centering
	\begin{tabular}{lccccc}
		\hline
		& Coefficient & Standard error & t & $P>|t|$ \\ 
		\hline
		Age  & -0.0571 & 0.031 & -1.840 & 0.070 \\
		Skin hydration    & 0.0132 & 0.023 & 0.574 & 0.567 \\
		Skin elasticity    & 0.0117 & 0.029 & 0.410 & 0.683  \\
		BMI & -0.0125 & 0.022 & -0.558 & 0.578 \\
		\hline
	\end{tabular}
	\caption{Regression model values for frequency range 20 to 50 Hertz.}
    \label{tab:RT1}
\end{table}
%%%%%%%%%%%%%%%%%%%%%%%%%%%%%%%%%%%%%%%%%%%%%%%%%%%%%%%%%%%%%%%%%%%%%%%%%%%%%%%%%%%%%%%%%%%%%%%%
\newpage 

\begin{table}[h!]
	\centering
	\begin{tabular}{lccccc}
		\hline
		& Coefficient & Standard error & t & $P>|t|$ \\ 
		\hline
		Age    & -0.0307 & 0.030 & -1.020 & 0.311 \\
		Skin hydration    & 0.0103 & 0.022 & 0.459 & 0.647 \\
		Skin elasticity    & 0.0161 & 0.028 & 0.582 & 0.562 \\
		BMI    & 0.0137 & 0.022 & 0.628 & 0.532 \\
		\hline
	\end{tabular}
	\caption{Regression model values for frequency range 50 to 110 Hertz.}
    \label{tab:RT2}
\end{table}

%%%%%%%%%%%%%%%%%%%%%%%%%%%%%%%%%%%%%%%%%%%%%%%%%%%%%%%%%%%%%%%%%%%%%%%%%%%%%%%%%%%%%%%%%%%%%%%
\newpage

\begin{table}[h!]
	\centering
	\begin{tabular}{lccccc}
		\hline
		& Coefficient & Standard error & t & $P>|t|$ \\ 
		\hline
		Age    & 0.0004 & 0.027 & 0.013 & 0.990 \\
		Skin hydration    & 0.0129 & 0.020 & 0.651 & 0.517 \\
		Skin elasticity    & 0.0162 & 0.025 & 0.662 & 0.510 \\
		BMI    & 0.0115 & 0.019 & 0.599 & 0.551 \\
		\hline
	\end{tabular}
	\caption{Regression model values for frequency range 110 to 230 Hertz.}
    \label{tab:RT3}
\end{table}

%%%%%%%%%%%%%%%%%%%%%%%%%%%%%%%%%%%%%%%%%%%%%%%%%%%%%%%%%%%%%%%%%%%%%%%%%%%%%%%%%%%%%%%%%%%%%%%
\newpage

\begin{table}[h!]
	\centering
	\begin{tabular}{lccccc}
		\hline
		& Coefficient & Standard error & t & $P>|t|$ \\ 
		\hline
		Age    & -0.0053 & 0.028 & -0.191 & 0.849 \\
		Skin hydration    & 0.0146 & 0.021 & 0.706 & 0.482 \\
		Skin elasticity    & 0.0228 & 0.026 & 0.888 & 0.377 \\
		BMI    & 0.0054 & 0.020 & 0.267 & 0.790 \\
		\hline
	\end{tabular}
	\caption{Regression model values for frequency range 230 to 450 Hertz.}
    \label{tab:RT4}
\end{table}

%%%%%%%%%%%%%%%%%%%%%%%%%%%%%%%%%%%%%%%%%%%%%%%%%%%%%%%%%%%%%%%%%%%%%%%%%%%%%%%%%%%%%%%%%%%%%%%%%%%%%%%%%%%%%%%%%%%%%%%%%%%%%%%%%
\section{Usage Notes}
Personalized classification models, such as {\em k} - medoids, trained individually for each subject, exhibit a geometric structure that remains insensitive to demographic variations. However, a critical gap remains in understanding how zero-shot or few-shot algorithms perform when tested on individuals from demographics not included in the training set. For instance, if a machine learning model is trained on data from young adults with low BMI, its ability to generalize to older adults with high BMI is not guaranteed. Likewise, developing efficient few-shot learning strategies to achieve accurate gesture classification for individuals outside the training demographics is essential.

Addressing these challenges is crucial for the successful deployment of devices like those described by \citeauthor{ctrl2024generic} Our dataset provides a valuable opportunity to investigate these issues, with the overarching goal of developing fair and inclusive algorithms that perform effectively across diverse individual profiles. Specifically, this dataset can serve as a benchmark for exploring efficient architecture designs for pretrained models, ensuring they learn generalizable features across individuals and enable zero-shot decoding for previously unseen subjects. Additionally, it offers a platform to examine the amount of data required from an individual for effective few-shot learning, helping to optimize model adaptation with minimal samples.
%%%%%%%%%%%%%%%%%%%%%%%%%%%%%%%%%%%%%%%%%%%%%%%%%%%%%%%%%%%%%%%%%%%%%%%%%%%%%%%%%%%%%%%%%%%%%%%%%%%%%%%%%%%%%%%%%%%%%%%%%%

\section*{Code availability}

\textcircled{\footnotesize 1} Codes are available at \href{https://github.com/HarshavardhanaTG/wristEMG}{https://github.com/HarshavardhanaTG/wristEMG}.

\section*{Acknowledgments}
 This work was supported by Meta Platforms Technologies (Facebook Research) with an award to L.M.M. through the Ethical Neurotechnology program and by the University of California Davis School of Medicine Cultivating Team Science Award to L.M.M. We would like to thank Stephanie Naufel at Facebook Reality Labs for her valuable guidance. 

H.T.G. is supported by Neuralstorm Fellowship, NSF NRT Award No. 2152260 and Ellis Fund administered by the University of California, Davis.

\section*{Author contributions}
H.T.G. contributed to the mathematical formulation, concept development, data analysis, experiment design, data collection software design, data collection, and manuscript preparation. N.K. contributed to data collection. C.C. contributed to data collection and experiment design. M.B. contributed to experiment design. S.A., S.K., S.L., Z.M., and F.R. contributed to data collection. S.S. contributed to data analysis. J.S. contributed to concept development, experiment design, and manuscript review. L.M.M. contributed to concept development, experiment design, and manuscript preparation.

\section*{Conflict of interest}
The authors declare no competing interests.

%%%%%%%%%%%%%%%%%%%%%%%%%%%%%%%%%%%%%%%%%%%%%%%%%%%%%%%%%%%%%%%%%%%%%%%%%%%%%%%%%%%%%%%%%%%%%%%%%%%%%%%%%%%%%%%%%%%%%%%%%%%%%%%%
%\bibliography{example_paper}
%\bibliographystyle{icml2025}

\end{document}